\documentclass[pra,superscriptaddress,floatfix,twocolumn,longbibliography]{revtex4-1}
\usepackage{graphicx}
\usepackage{wasysym}
\usepackage{color}

\def\lsco{La$_{2-x}$Sr$_x$CuO$_4$}
\def\lbco{La$_{2-x}$Ba$_x$CuO$_4$}

\def\ybco{YBa$_2$Cu$_3$O$_{6+x}$}

\def\bscco{Bi$_2$Sr$_2$CaCu$_2$O$_{8+\delta}$}

\def\och{O$_{\rm ch}$}
\def\cuch{Cu$_{\rm ch}$}
\def\oap{O$_{\rm ap}$}
\def\oplone{O$_{\rm pl}$(1)}
\def\opltwo{O$_{\rm pl}$(2)}
\def\opl{O$_{\rm pl}$}
\def\cupl{Cu$_{\rm pl}$}

\newfont{\fc}{cmssbx10 scaled 1000}

\begin{document}

\title{Charge-screening role of $c$-axis atomic displacements in YBa$_2$Cu$_3$O$_{6+x}$ and related superconductors}

\author{E. S. Bo\v{z}in}
\affiliation{Condensed Matter Physics \&\ Materials Science Department, Brookhaven National Laboratory, Upton, NY 11973-5000, USA}
\author{A. Huq}
\affiliation{Chemical and Engineering Materials Division, Oak Ridge National Laboratory, Oak Ridge, TN 37831, USA}
\author{Bing Shen}
\author{H. Claus}
\author{W. K. Kwok}
\affiliation{Materials Science Division, Argonne National Laboratory, Argonne, Illinois 60439, USA}
\author{J. M. Tranquada}
\affiliation{Condensed Matter Physics \&\ Materials Science Department, Brookhaven National Laboratory, Upton, NY 11973-5000, USA}
\date{\today}
\begin{abstract}
The importance of charge reservoir layers for supplying holes to the CuO$_2$ planes of cuprate superconductors has long been recognized.  Less attention has been paid to the screening of the charge transfer by the intervening ionic layers.  We address this issue in the case of \ybco, where CuO chains supply the holes for the planes.  We present a simple dielectric-screening model that gives a linear correlation between the relative displacements of ions along the $c$ axis, determined by neutron powder diffraction, and the hole density of the planes.  Applying this model to the temperature dependent shifts of ions along the $c$ axis, we infer a charge transfer of 5--10\%\ of the hole density from the planes to the chains on warming from the superconducting transition to room temperature.  Given the significant coupling of $c$-axis displacements to the average charge density, we point out the relevance of local displacements for screening charge modulations and note recent evidence for dynamic screening of in-plane quasiparticles.  This line of argument leads us to a simple model for atomic displacements and charge modulation that is consistent with images from scanning-tunneling microscopy for underdoped \bscco.
\end{abstract}
%
\maketitle

\section{Introduction}

The interactions key to electron pairing and superconductivity in the cuprates occur within the CuO$_2$ planes, and it follows that most research has focused on characterization and analysis of in-plane behaviors \cite{kast98,dama03,baso05,norm05,lee06,scal12a,frad15}.  There has been some consideration of the impact of out-of-plane disorder on the scattering rate of in-plane quasiparticles \cite{varm01} and on the superconducting transition temperature, $T_c$ \cite{eisa04,hobo09}.   In a more positive sense, there has been recognition that screening of long-range Coulomb interactions within the planes can benefit from responses outside of the planes \cite{kive02b,ragh12}.  In particular, the beneficial impact of charge reservoir layers on the screening of long-range Coulomb interactions in the CuO$_2$ planes has recently been emphasized \cite{ragh12}.   The purpose of the present paper is to investigate the role of the intervening ions in the screening process.

In the case of \ybco (YBCO), the role of the charge reservoir layers is played by the ``chain'' layers \cite{jorg90,cava90}.   As demonstrated by x-ray spectroscopic studies \cite{tran88b,tole92,uimi92}, adding oxygen to the chain layers converts Cu$^{1+}$ ions to Cu$^{2+}$ plus a hole.  A fraction of that hole is transferred to the neighboring planes, with dielectric screening provided by adjustments to the spacing between the intervening Ba$^{2+}$ and O$^{2-}$ layers \cite{tran90}.  While the systematic trends of the atomic shifts with doping are clear from the early neutron diffraction studies of structure with doping \cite{jorg90,cava90}, we are not aware of any quantitative analysis of those results.

In this paper, we present a model that connects the hole density $p$ per planar Cu to the atomic shifts of ions along the $c$ axis relative to their positions in YBa$_2$Cu$_3$O$_6$.  Applying this model to structural parameters from neutron powder diffraction, good agreement is obtained with the $p$ vs.\ $x$ data of Liang {\it et al.} \cite{lian06} with a single multiplicative scaling factor of order one.  From new neutron powder diffraction data, we find that atomic positions along the $c$ axis shift with temperature, from which we infer a decrease in $p$ of 0.01 (10\%\ of $p$ for $x=0.56$) on warming from $T_c$ to 300~K.  

Besides our observation of thermal shifts in average dielectric screening, there is also evidence from  electron-energy-loss spectroscopy (EELS) studies of \bscco\ for dynamic coupling of in-plane electrons to $c$-axis optical phonons \cite{qin10,vig15,phel93}, especially those involving apical and in-plane oxygens \cite{kova04}.  We discuss the relevance of such modes to the short-range charge-density-wave (CDW) order detected in \ybco\ and related cuprates \cite{ghir12,chan12a,comi15c}, where a recent analysis of superlattice intensities indicates significant atomic displacements along the $c$ axis \cite{forg15}.   In high magnetic fields and temperatures below $T_c$, a new CDW, modulated only along the $b$ axis, develops \cite{wu11,gerb15,chan15} that is compatible with the anisotropic soft bond-bending phonon observed by inelastic neutron scattering \cite{raic11b}.  Assuming the latter displacement pattern is also relevant in \bscco, we obtain a model for the hole distribution in good agreement with images from scanning tunneling microscopy (STM) studies \cite{kohs07}.

The rest of the paper is organized as follows.  In the following section, we briefly describe our sample preparation and neutron powder diffraction measurements.  In Sec.~\ref{sc:results}, we present sample characterizations, describe the charge-transfer screening model, and use it analyze the changes with doping and temperature.  We also note an anomalous enhancement of the orthorhombic strain due to differing lattice stiffnesses along the $a$ and $b$ axes.  In Sec.~\ref{sc:disc}, we discuss the role of $c$-axis atomic displacements and their relevance to recent studies of CDW order in several cuprates.  The paper concludes with a summary in Sec.~\ref{sc:sum}.

\section{Experimental methods}

The synthesis of \ybco\ powders followed the procedures established in \cite{jorg90,veal90}. Specifically, the samples were prepared from powders of Y$_2$O$_3$ (99.999\% Sigma-Aldrich) BaCO$_3$ (99.999\% Sigma-Aldrich) and CuO (99.999\% Sigma-Aldrich). These powders were mixed stoichiometrically and ground in an agate mortar for one hour. Subsequently, the mixtures were placed in a BaZrO$_3$ crucible, heated to 960 $^\circ$C, and sintered for 24~h in flowing O$_2$ (99.99\%). Following that, the mixture was cooled and reground for 1 h, and then pressed into pellets with a diameter and thickness of 2 cm and 2.5 mm, respectively. The pellets were reheated to 960 $^\circ$C in flowing O2(99.99\%) for 24 h, followed by quenching in liquid nitrogen.

\begin{figure}[t]
\begin{center}
\includegraphics[width=0.8\columnwidth]{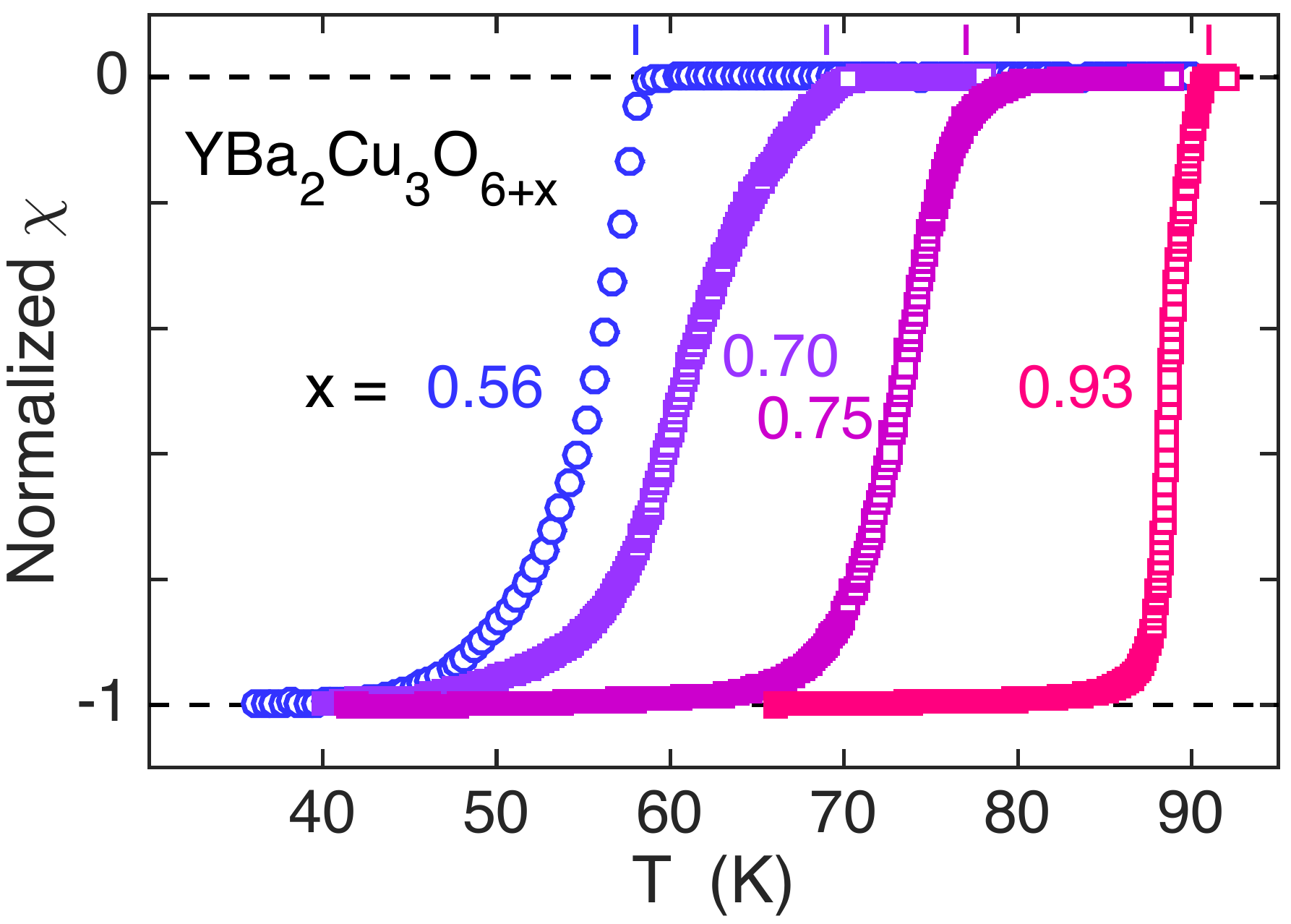}
\caption{\label{fg:chi} 
(Color online)  Normalized bulk susceptibility measured on warming in a 1 Oe field after zero-field cooling.  Vertical bars at the top indicate the superconducting transitions.}
\end{center}
\end{figure}

\begin{table}[b]
\caption{\label{tab:S} Annealing conditions used for each sample of \ybco\ as characterized by its measured values of $x$ and $T_c$.}
\begin{ruledtabular}
\begin{tabular}{ccccc}
$x$  & $T_c$ & gas mixture & $T_{\rm anneal}$ & Time \\
        & (K) & & ($^\circ$C) & (days) \\
\hline
0.56 & 58 & 0.1\%\, O$_2$ + 99.9\%\,Ar & 450 & 14\\
0.70 & 69 & 1\%\, O$_2$ + 99\%\,Ar & 430 & 10\\
0.75 & 77 & 1\%\,O$_2$ + 99\%\,Ar &400 & 14\\
0.93 & 91 & 100\%\,O$_2$ & 440 & 10\\
\end{tabular}
\end{ruledtabular}
\end{table}

The oxygen content of the polycrystalline samples was adjusted by annealing in an appropriate partial pressure of O$_2$, followed by quenching in liquid nitrogen.  The superconducting transition temperature, $T_c$, was determined from the zero-field-cooled susceptibility data shown in Fig.~\ref{fg:chi}.  To define $T_c$, we drew a line tangent to $\chi(T)=-0.15$ and found its intersection with $\chi=0$.    The oxygen content $x$ was determined from the analysis of the neutron powder diffraction data, as described below.  The samples and annealing conditions are listed in Table~\ref{tab:S}.

\begin{figure}[t]
\begin{center}
\includegraphics[width=\columnwidth]{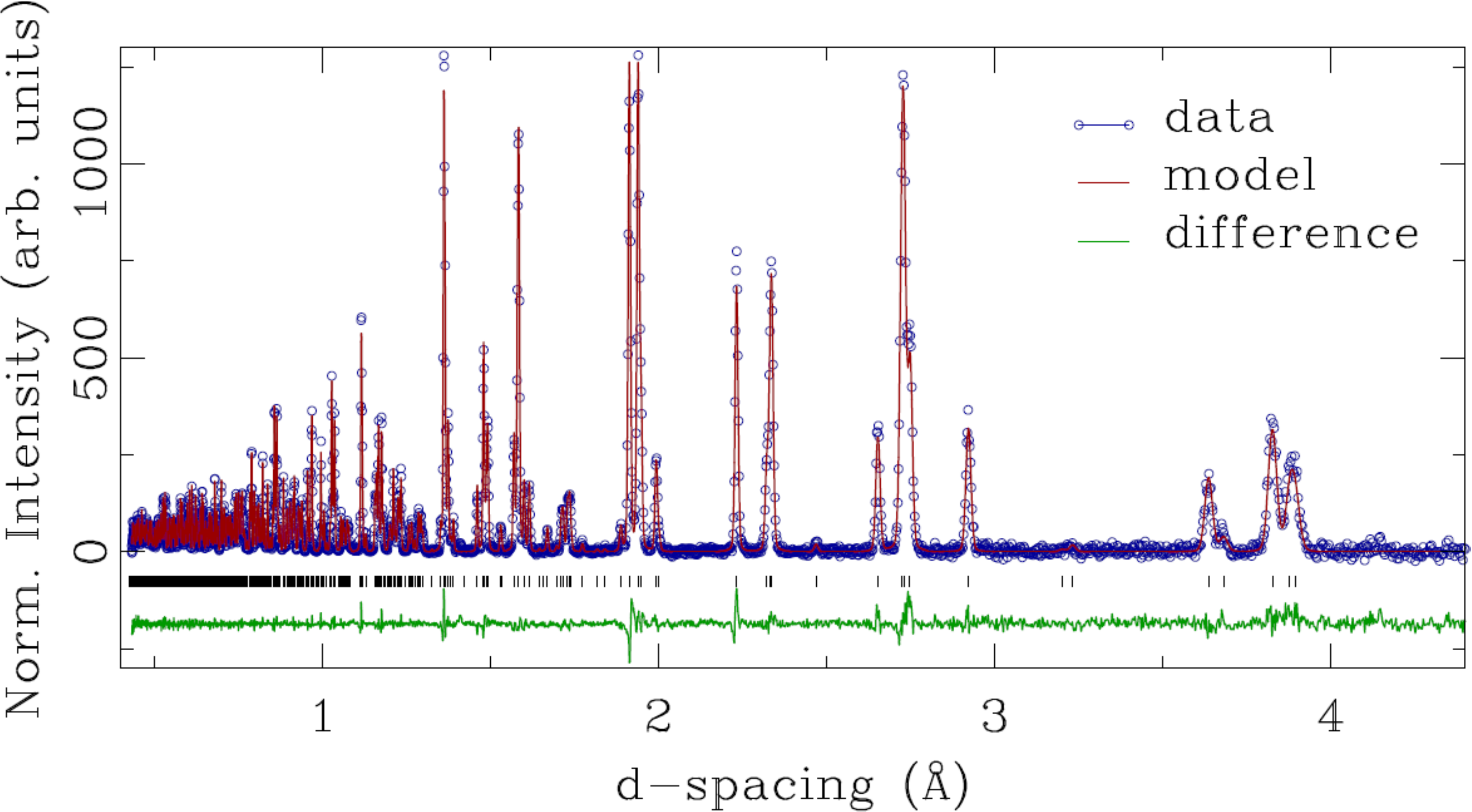}
\caption{\label{fg:refine} 
(Color online)  Typical Rietveld refinement of the $Pmmm$ structural model (line) to the $x=0.56$, $T_c=58$~K sample data (circles) at 10~K.  Vertical bars indicate calculated peak positions; line at bottom indicates difference between data and model.  Goodness of fit parameters are $\chi^2=3.59$, $R_{\rm wp}=4.99\%$, and Bragg factor $R=6.86\%$.}
\end{center}
\end{figure}

Neutron powder diffraction measurements were performed on the POWGEN instrument, BL-11A  at the Spallation Neutron Source, Oak Ridge National Laboratory.  The samples were cooled in a closed-cycle refrigerator with {\it in situ} sample changer.  The average structure was assessed through Rietveld refinements \cite{riet67} to the raw diffraction data using GSAS \cite{lars87} operated under EXPGUI \cite{toby01}, utilizing orthorhombic space group $Pmmm$.  No diffraction peaks from impurity phases were observed, from which we estimate that the volume fraction of impurities is $<1$\%.  A typical refinement is presented in Fig.~\ref{fg:refine}.  Isotropic mean square displacements from the fits exhibit a smooth and conventional increase with temperature, with no obvious anomalies.  Relative position coordinates $z$ along the $c$ axis are consistent with those of Cava {\it et al.} \cite{cava90}, generally to better than 0.001.  

\section{Results}
\label{sc:results}

\subsection{Sample characterization}

\begin{figure}[t]
\begin{center}
\includegraphics[width=0.9\columnwidth]{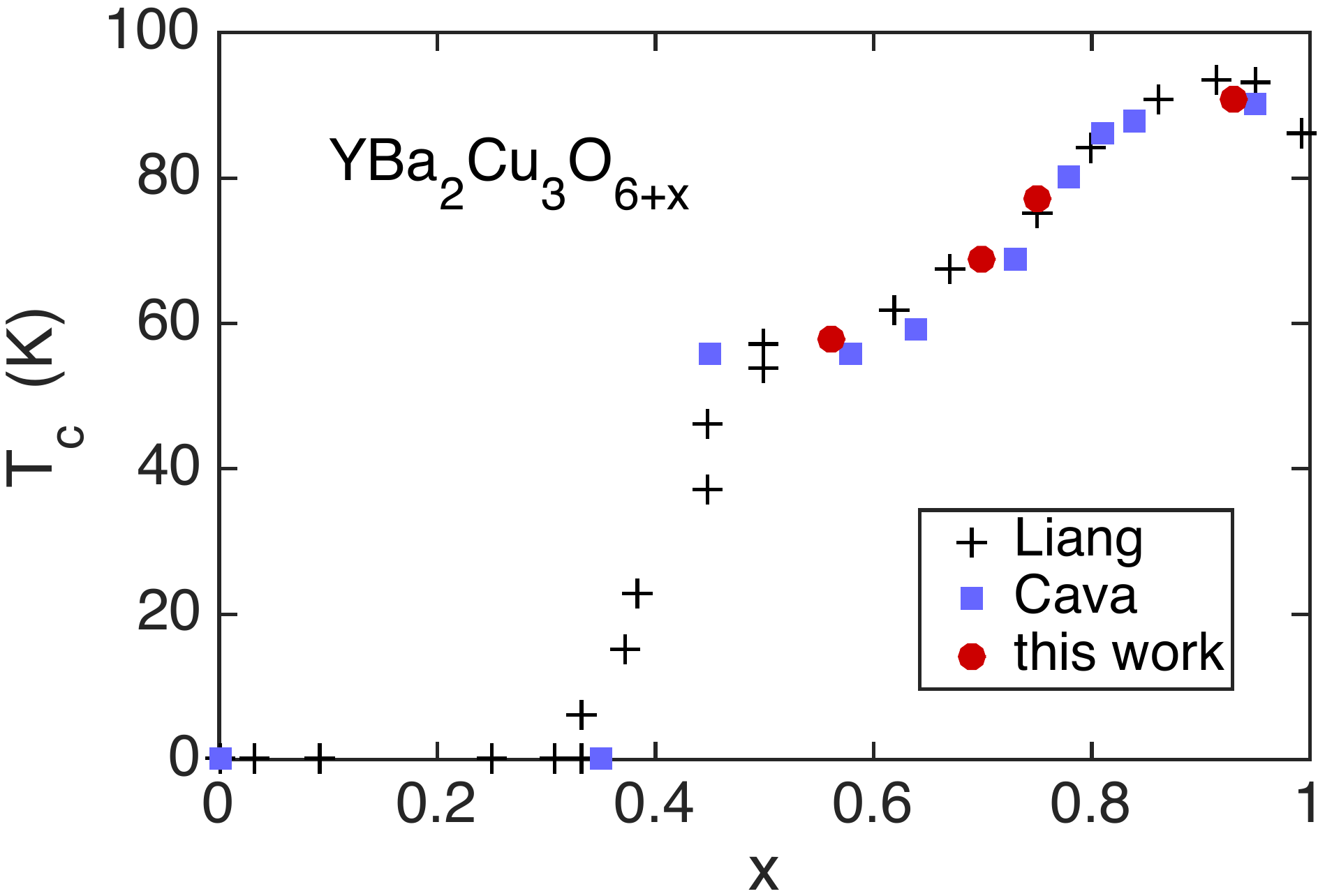}
\caption{\label{fg:tcvx} 
(Color online)  Superconducting transition temperature vs.\ oxygen content in \ybco: data from Liang {\it et al.} \cite{lian06} (crosses), Cava {\it et al.} \cite{cava90} (squares), and present work (circles).}
\end{center}
\end{figure}

\begin{figure}[b]
\begin{center}
\includegraphics[width=0.9\columnwidth]{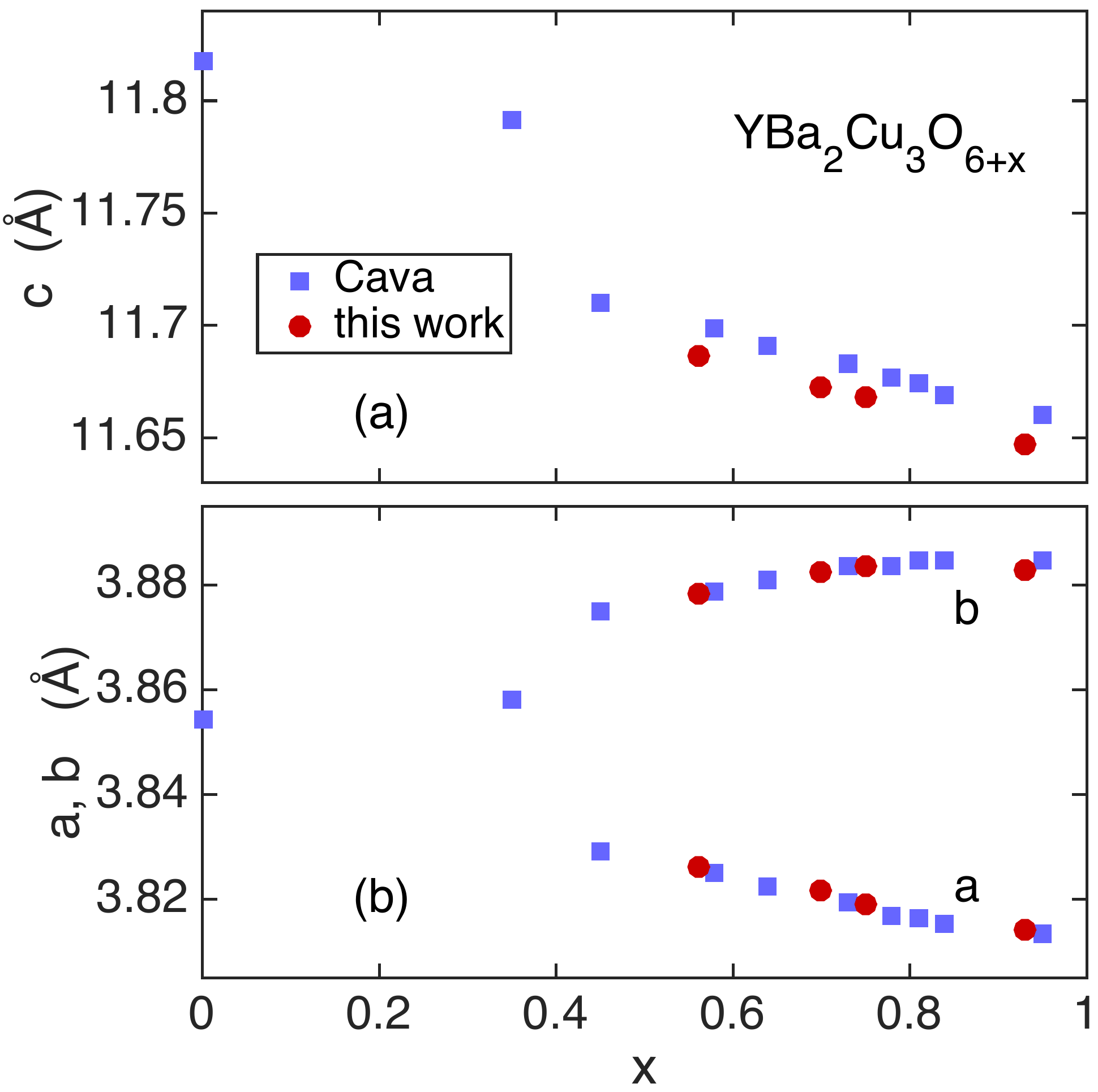}
\caption{\label{fg:abcvx} 
(Color online)  Lattice parameters in \ybco: (a) $c$ and (b) $a$ and $b$.  Data from Cava {\it et al.} \cite{cava90} at $T=5$~K (squares) and present work at 10~K (circles).
}
\end{center}
\end{figure}

To compare the present samples of \ybco\ with previous work, we plot $T_c$ vs.\ $x$ in Fig.~\ref{fg:tcvx}.  We find good agreement with both the early powder study of Cava {\it et al.} \cite{cava90} and the single-crystal results of Liang {\it et al.} \cite{lian06}.  The doping dependence of the low-temperature lattice parameters is plotted in Fig.~\ref{fg:abcvx}.  Again, we see good agreement with Cava {\it et al.} \cite{cava90} for $a$ and $b$; however, our $c$ lattice parameters are systematically smaller, which might be a consequence of different annealing conditions as these are known to impact the degree of Cu-O chain ordering.

\subsection{Charge transfer from chains to planes}

The superconducting properties of \ybco\ are dominated by the planes; however, the carrier concentration is controlled by the density of oxygen ions in the chain layers.  A fraction of the holes created in the chain layers are transferred to the planes, and dielectric screening of this charge transfer occurs through displacements of ions along the $c$ axis.  A consequence is that there is a close connection between planar carrier concentration and relative coordinates along the $c$ axis.

\begin{figure}[t]
\begin{center}
\includegraphics[width=\columnwidth]{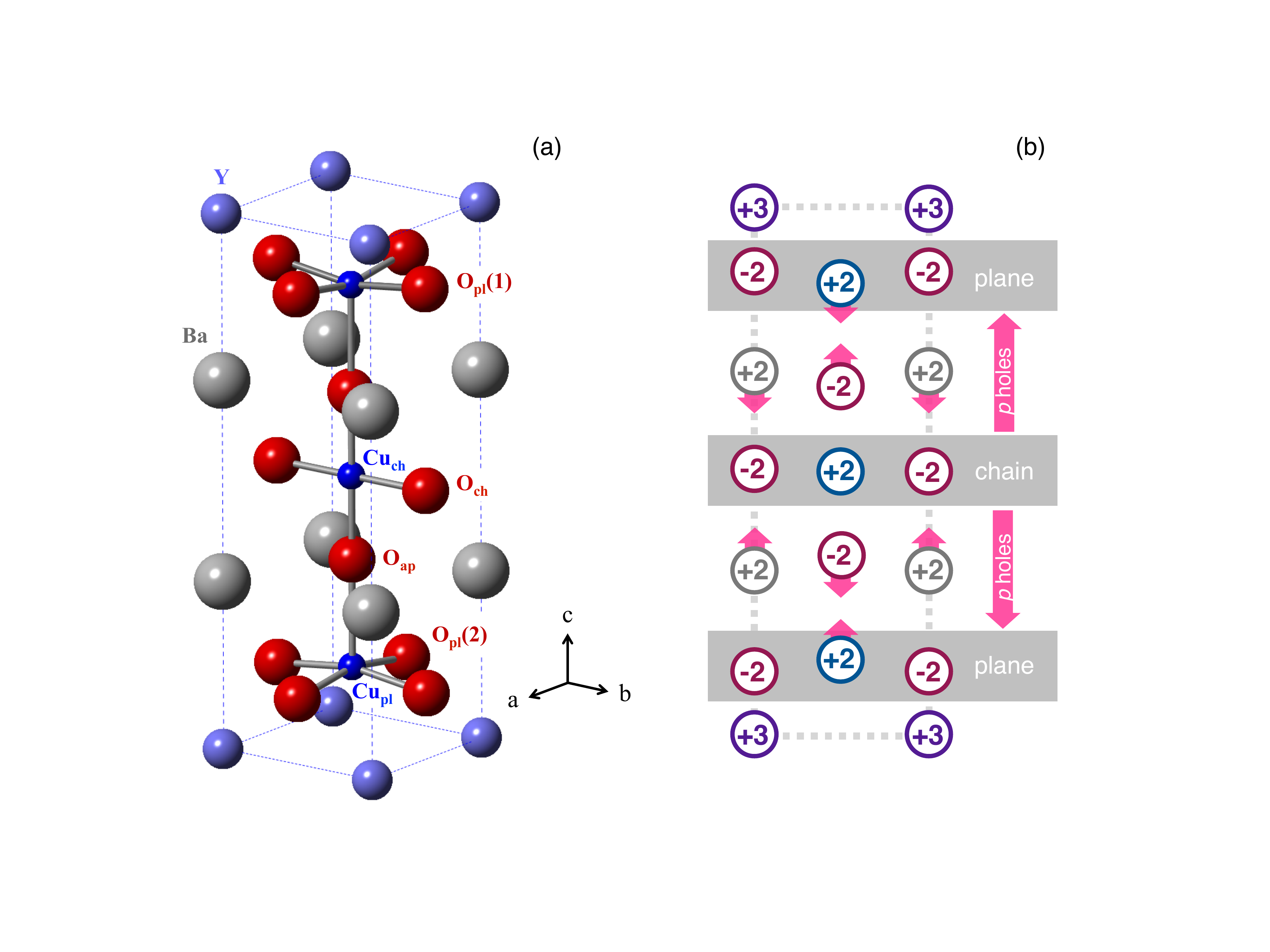}
\caption{\label{fg:uc} (Color online)
(a) Diagram of the orthorhombic unit cell of \ybco, with the distinct sites labelled. Note that there are mirror planes at the Y and the chain layers.  We choose to place the origin of $z$, the relative coordinate along $c$, at the chain layer.  (b) Schematic of the unit cell projected along [100], labelled by nominal ionic valences.  Long arrows indicate the transfer of doped holes from the Cu-O chains to the CuO$_2$ planes; short arrows indicate the compensating atomic displacements.  Dashed line outlines the unit cell.}
\end{center}
\end{figure}

To analyze the hole creation, transfer, and screening, we start with some commonly accepted results.  The doped holes have a predominant O $2p_\sigma$ character, as determined by measurements at O $K$ and Cu $L_3$ absorption edges \cite{romb90,chen91,nuck95}.  The CuO$_2$ planes have a net negative charge, so that adding holes (removing electrons) tends to cause a decrease in the in-plane Cu-O bond length, as observed in diffraction studies of single-layer cuprates \cite{rada94}.  \ybco, whose crystal structure is indicated in Fig.~\ref{fg:uc}(a), is somewhat unique in that the doping involves inserting O atoms into the \och\ lattice sites that are completely empty when $x=0$, at which point the 2-fold coordinated \cuch\ sites have a valence of $+1$ and the CuO$_2$ planes are undoped ({\it i.e.}, hole concentration $p$ per \cupl\ is zero).  Initally, inserting a neutral O between a pair of \cuch\ sites (making them 3-fold coordinated) pulls an electron from each, resulting in a pair of Cu$^{2+}$ ions and O$^{2-}$, while maintaining $p=0$.  With further O addition, chain segments will form, leading to a transformation from tetragonal to orthorhombic symmetry, with the chains oriented along the $b$ axis.  For an infinite chain of 4-fold coordinated \cuch\ sites, each added O creates one hole, in addition to a Cu$^{2+}$ \cite{tran88b,tole92,uimi92}.   To obtain superconductivity, some of these holes need to be transferred from the \och\ sites in the chains to the \opl\ sites in the planes.  In between these positions, we have the Ba$^{2+}$-O$_{\rm ap}^{2-}$ layer, with the Ba and \oap\ atoms having slightly different relative heights $z$ along the $c$ axis, and the \cupl\ sites that have a different $z$ from the \opl\ sites.  We then expect that the charge transfer from the chains to the planes is screened by displacements of \oap\ relative to Ba, and \cupl\ relative to \opl, as indicated schematically in Fig.~\ref{fg:uc}(b).  In fact, a large jump in the Ba-\oap\ spacing occurs at the tetragonal to orthorhombic transition \cite{cava90}, where there is a concomitant jump in $p$ from negligible to finite \cite{lian06}.   Note that the insertion of the O atoms into the chains causes the $b$ lattice parameter to expand, competing with the tendency for hole doping to reduce the \cupl-\opl\ bond length.

To describe this behavior, we will consider a rigid ion model, assuming integer valences of the ions.  (This is not entirely accurate, as there is experimental evidence for a small hole concentration on the \oap\ sites \cite{nuck95}; however, we will ignore such effects for the sake of simplicity.)  We take the $x=0$ composition as the reference configuration, and consider a two-step process in which we first introduce neutral O atoms into the chains and then transfer a density of holes $p$ from the chain layer to each neighboring CuO$_2$ plane.   This creates a potential difference $V_1$, given by
\begin{equation}
 V_1 = p(z_{\rm O_{pl}} - z_{\rm O_{ch}})c / \epsilon_0,
\end{equation}
where $z_jc$ is the average distance from the chain layer along the $c$ axis for atoms at site $j$, evaluated for the corresponding value of oxygen content $x$, and $\epsilon_0$ is the vacuum permittivity. (Note that $z_{\rm O_{ch}}=0$.)  This potential difference will be countered by displacements of the intervening ions [O$_{\rm ap}^{2-}$, Ba$^{2+}$, Cu$_{\rm pl}^{2+}$, as indicated in Fig.~\ref{fg:uc}(b)], resulting in $V_2$, given by
\begin{equation}
  V_2 = 2[d(x)-d(0)]/\epsilon_0,
\end{equation} 
where
\begin{equation}
  d(x) = (z_{\rm Ba}-z_{\rm O_{ap}} + z_{\rm Cu_{pl}}-z_{\rm O_{pl}})c
  \label{eq:d}
\end{equation}
and we take $z_{\rm O_{pl}}$ to be the average of the $z$ parameters for the \oplone\ and \opltwo\ sites.  Assuming that $V_1+V_2=0$, we get
\begin{equation}
  p = -2\alpha {d(x)-d(0) \over (z_{\rm O_{pl}} - z_{\rm O_{ch}})c},
  \label{eq:p}
\end{equation}
where $\alpha$ is a doping-independent correction factor of order 1 that we include to account for corrections to our simple  model.   

\begin{figure}[t]
\begin{center}
\includegraphics[width=0.9\columnwidth]{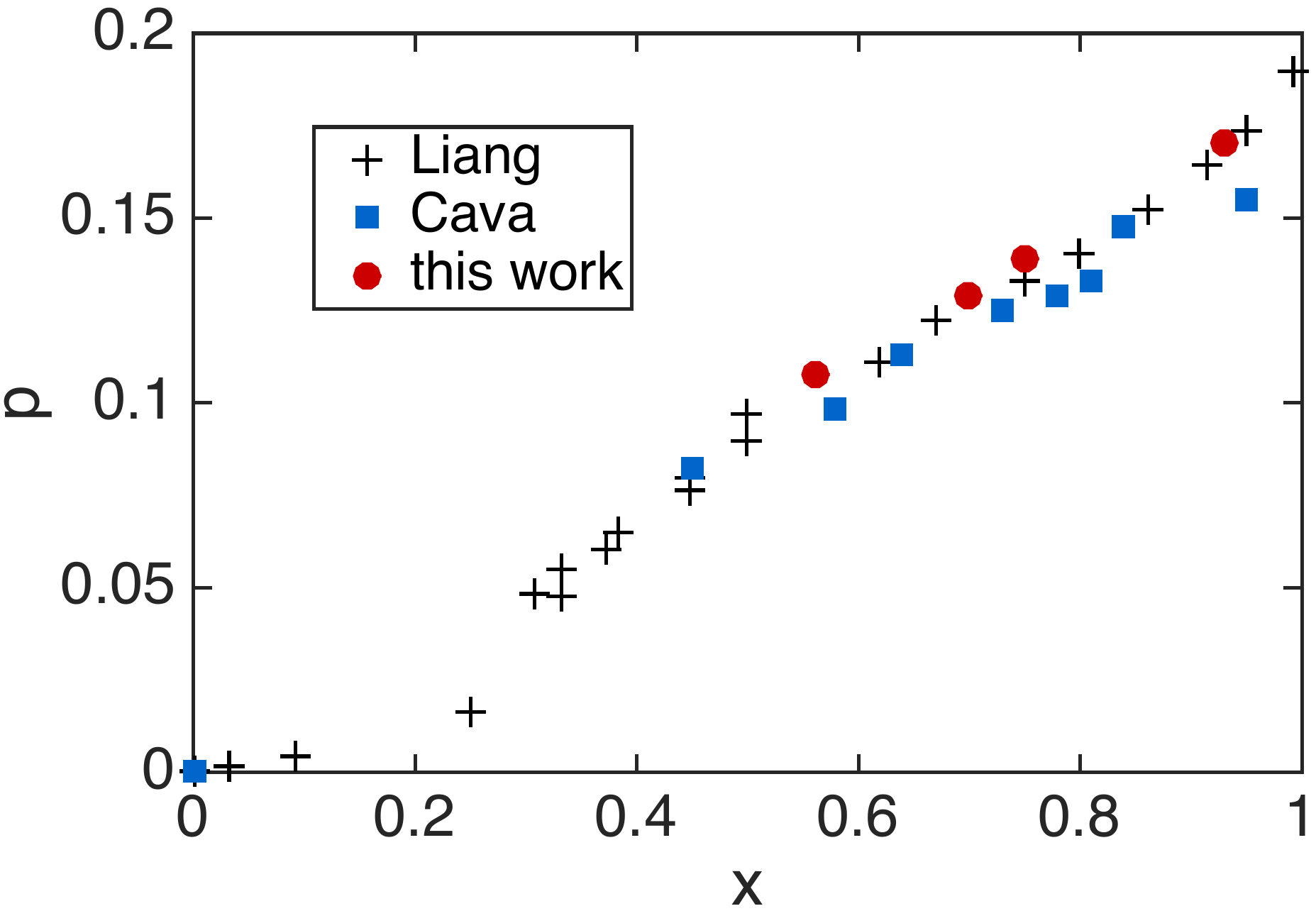}
\caption{\label{fg:pvx} (Color online)
Plot of $p$ vs.\ $x$ in \ybco\ from Liang {\it et al.} \cite{lian06} (crosses) and calculated, as discussed in the text, from structural parameters determined by Cava {\it et al.} \cite{cava90} (squares) and in the present work (circles).}
\end{center}
\end{figure}

In order to test this model, we need data for $p$ vs.\ $x$.  It has been challenging to directly determine $p$ in \ybco.  Perhaps the best analysis of the hole concentration is that by Liang {\it et al.} \cite{lian06}; their results for $p$ vs.\ $x$ are plotted in Fig.~\ref{fg:pvx}.   We have also calculated $p$ from the low-temperature structural parameters determined by Rietveld refinement in the present work and by Cava {\it et al.} \cite{cava90} using Eqs.~(\ref{eq:d}) and (\ref{eq:p}).  (Note that we must reference our own results to Cava's $x=0.0$ sample.)  Fixing $\alpha=1.4$, we obtain the results shown in Fig.~\ref{fg:pvx}.   The differences between $p$ values at similar $x$ values between our results and those of Cava {\it et al.} \cite{cava90} are consistent with the differences in the $c$ lattice parameters in Fig.~\ref{fg:abcvx} and the relationship between $p$ and $c$ determined by Liang {\it et al.} \cite{lian06}.

\subsection{Temperature dependence of the charge transfer}

\begin{figure}[t]
\begin{center}
\includegraphics[width=0.9\columnwidth]{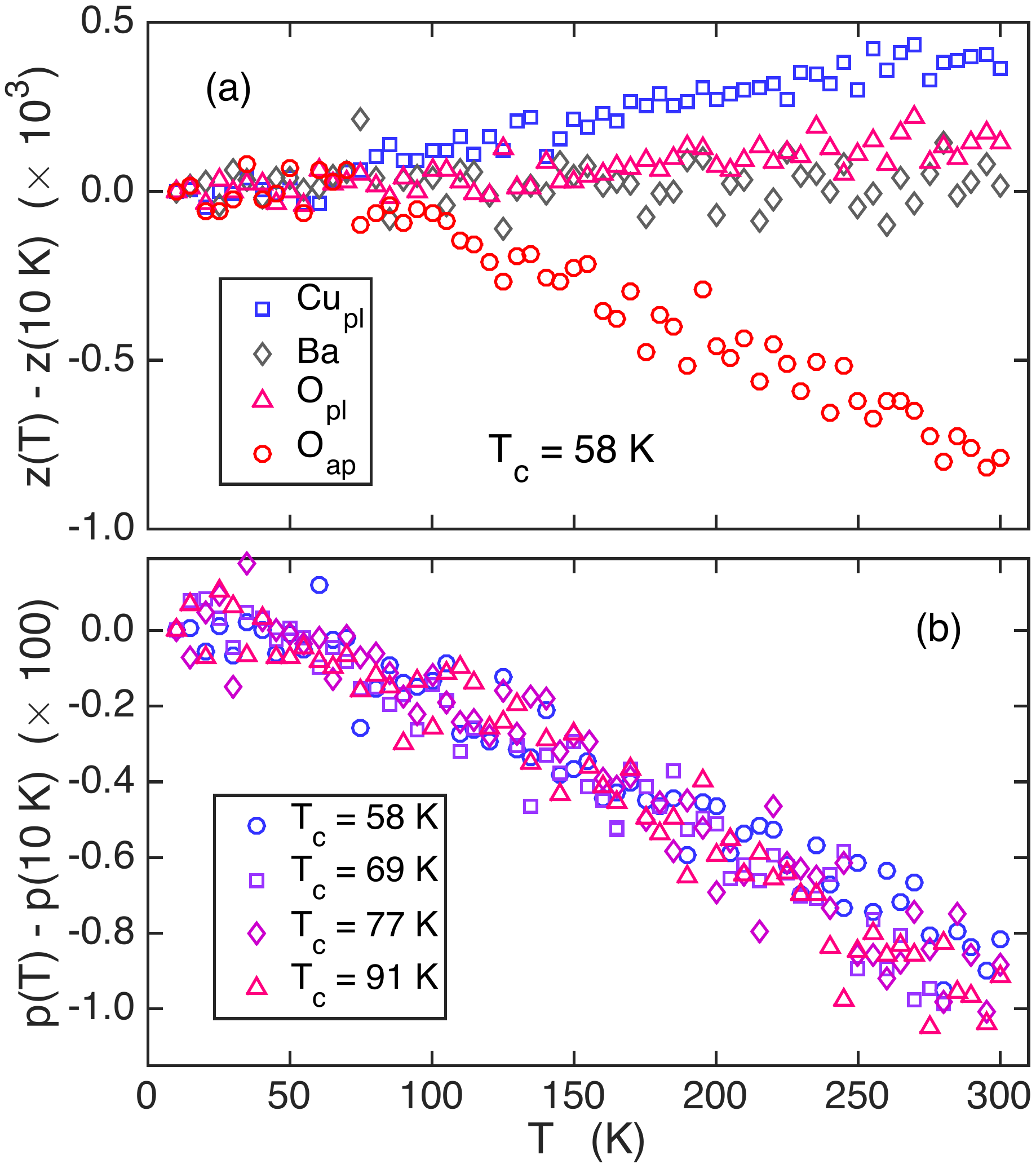}
\caption{\label{fg:dz} (Color online)
Relative temperature dependence of the $z$ parameters for the \cupl (squares), Ba (diamonds), \opl (triangles), and \oap\ (circles) sites determined by Rietveld refinement for the $x=0.56$ sample.  Uncertainties are comparable to the fluctuations.}
\end{center}
\end{figure}

Our neutron powder diffraction measurements were performed in small temperature steps from 10 to 300~K.  From the Rietveld refinements of these data, we obtain the temperature dependence of the structural parameters.  In Fig.~\ref{fg:dz}(a), we have plotted the temperature dependences of the $z$ parameters relative to 10-K values.  The biggest thermal changes occur for the \cupl\ and \oap\ sites.  Given that these ions provide screening of the charge transfer from the chains to the planes, the shifts in relative positions imply a change in the average hole concentration $p$ of the CuO$_2$ planes.  Using Eqs.~(\ref{eq:d}) and (\ref{eq:p}), we have calculated the temperature dependence of $p$ for each of our samples.  To emphasize the similarities, Fig.~\ref{fg:dz}(b) shows the change in $p$ relative to its value at 10~K.  Each sample shows a decrease in $p$ of $\sim0.01$ on warming to 300~K.  For the $T_c=58$~K sample, this corresponds to 10\%\ of $p$.

It would be useful to compare our inferred $T$-dependent change in $p$ with other measurements of $p(T)$; however, this proves to be a challenge.  One might hope to get $p$ from measurements of the Hall coefficient, but the Hall coefficient is temperature dependent due to correlation effects, and hence is not a precise measure of $p$ \cite{ando04,gork06}.  Nevertheless, we note that a recent x-ray spectroscopy study of YBa$_2$Cu$_3$O$_{6.9}$ \cite{magn14} is suggestive of a small decrease in $p$ when the temperature is raised, in qualitative agreement with our analysis.

\subsection{Temperature dependence of lattice parameters and orthorhombic strain}

The temperature dependence of lattice parameters can be sensitive to electronic transitions, as, for example, demonstrated in the case of charge-stripe ordering in La$_{1.67}$Sr$_{0.33}$NiO$_4$ \cite{abey13}.   Given the variety of unusual transitions reported for YBCO, it is worthwhile to briefly examine the data.  

\begin{figure}[t]
\begin{center}
\includegraphics[width=0.9\columnwidth]{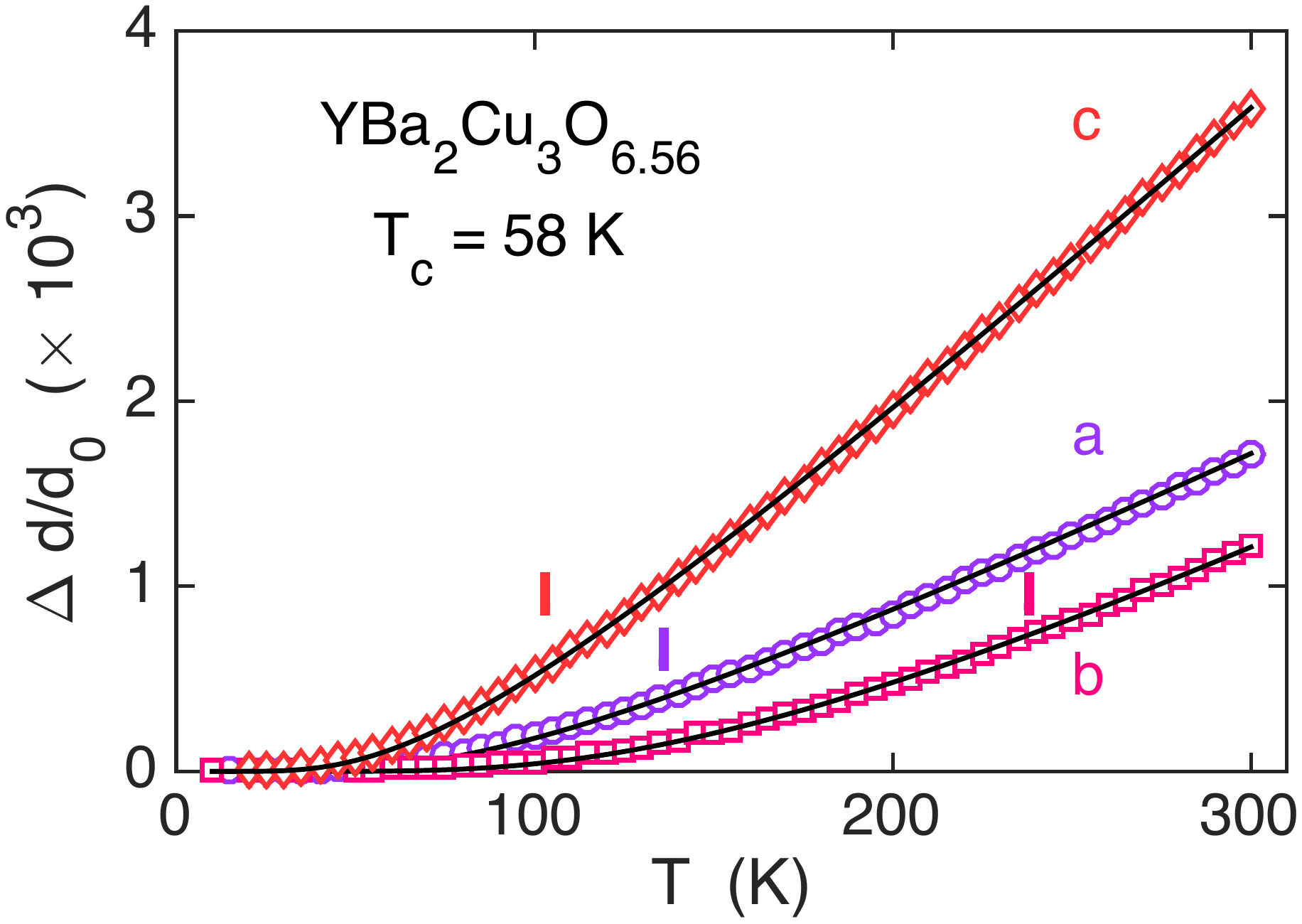}
\caption{\label{fg:abc} (Color online)
Temperature dependence of the normalized change in the lattice parameters $a$, $b$, and $c$.  Lines through the data points are fits with Eqs.~(\ref{eq:dd}) and (\ref{eq:f}), where the respective values of the parameter $T_0$ are indicated by the vertical bars.}
\end{center}
\end{figure}

Figure~\ref{fg:abc} shows the temperature dependence of $\Delta d(T)/d_0$ for the $x=0.56$ sample, where $d_0=d(10 K)$, $\Delta d(T) = d(T)-d_0$, and $d=a$, $b$, $c$.  We note that the temperature dependence of each lattice parameter is similar in form to that of the mean-square relative displacement  (MSRD) of a nearest-neighbor bond, which is often parametrized well by the Einstein model, $A\coth(T_0/T)$, with $T_0$ proportional to the Einstein frequency for the nearest neighbor bond \cite{knap85}.  While thermal expansion is due to anharmonicity and the MSRD formula is for harmonic behavior, the anharmonic effects will grow as the MSRD increases.   This has motivated the use of a one-parameter empirical model corresponding to:
\begin{equation}
  {\Delta d(T) \over d_0}   = {\Delta d(300~K) \over d_0}{f(T) \over f(300~K)},   
  \label{eq:dd}
\end{equation}
where
\begin{equation}
  f(T) = \coth\left({T_0/ T}\right) - 1.
  \label{eq:f}
\end{equation}
The lines through the data points are fits to this model with a different $T_0$ for each axis.  For example, $T_0$ is 136(6), 238(12), and 103(4)~K for $a$, $b$, and $c$, respectively, for the $x=0.56$ sample.  For $x=0.93$, there are increases of $T_0$ by 5\%, 1\%, and 12\%\ for $a$, $b$, and $c$, respectively.

One observation is that there are no obvious anomalies in the temperature dependences of the lattice parameters.  There is a similar smooth evolution of each lattice parameter, becoming approximately linear in temperature above $T_0$.  The magnitudes of $T_0$ appear to reflect the relative lattice stiffnesses along the corresponding directions.  The bonding along the $c$ axis is relatively ionic, resulting in weak interatomic forces and the smallest $T_0$.  The largest $T_0$ occurs for the $b$ axis, which is stiffer than $a$ because of the \cuch--\och\ bonds.

\begin{figure}[t]
\begin{center}
\includegraphics[width=0.9\columnwidth]{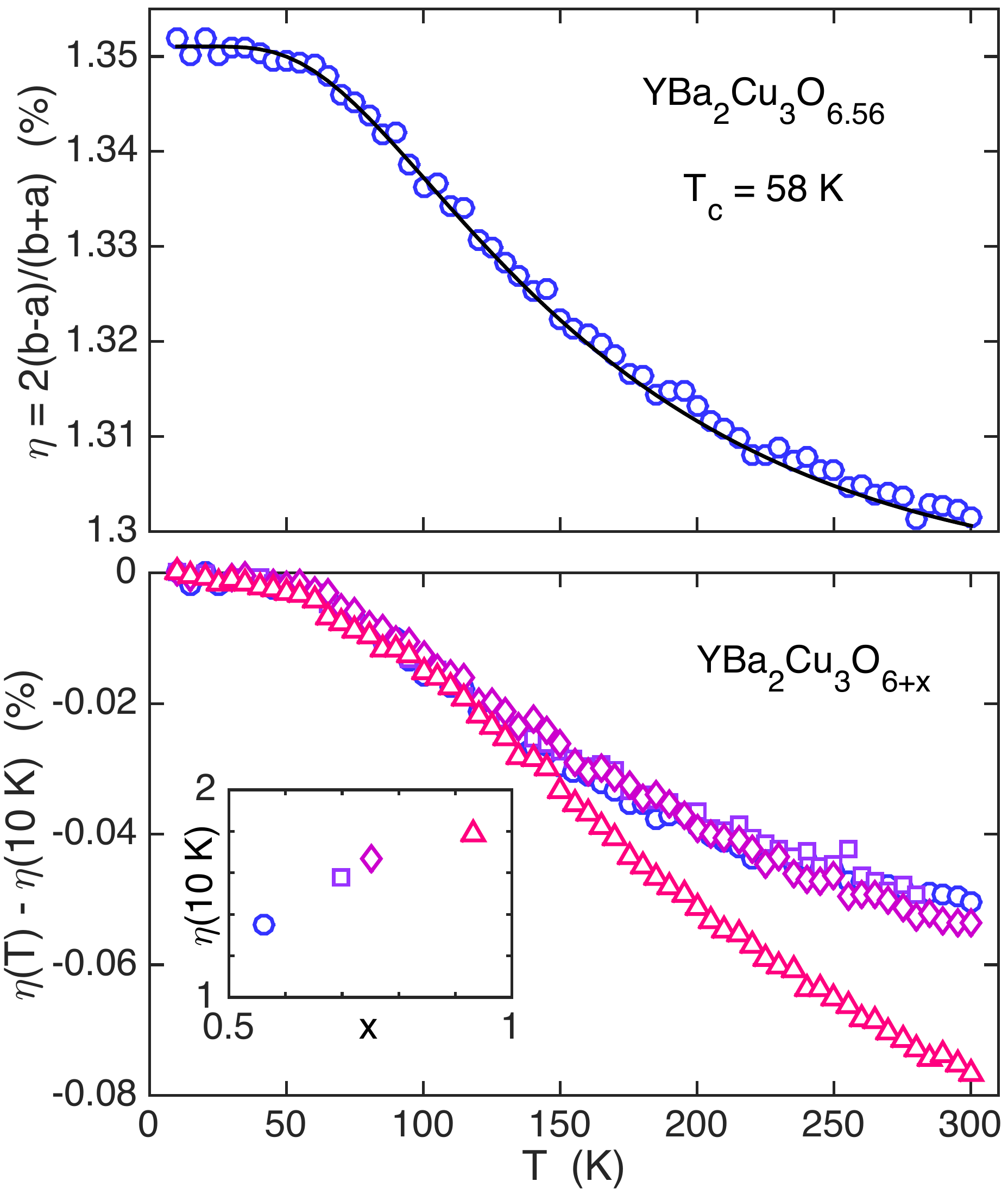}
\caption{\label{fg:s6} (Color online)
(a) Orthorhombic strain vs.\ temperature for the $x=0.56$, $T_c=58$~K sample. Line is calculated from the fits to $a(T)$ and $b(T)$, shown in Fig.~\ref{fg:abc}.  (b) Temperature dependence of the strain relative to the strain at 10~K for all four samples.  Inset shows the values of the low temperature strain values as a function of $x$, providing an implicit symbol legend.}
\end{center}
\end{figure}

The orthorhombic strain, $\eta = 2(b-a)/(b+a)$, is of interest because electronic nematicity should couple to it \cite{daou10,nie14}.  In Fig.~\ref{fg:s6}(a), we plot the strain obtained experimentally for the $x=0.56$ sample.  There appears to be an anomalous enhancement of the magnitude of the temperature derivative of the strain on cooling below 200~K, which is the range where nematic effects appear in measurements of the Nernst effect \cite{daou10}.   
Could this enhanced rate of growth of the strain be due to coupling to conduction electrons?  While we cannot rule out such a coupling, a more mundane effect provides an adequate explanation of the observations.
The ``anomaly'' in the strain growth appears to be a simple consequence of the  different values of $T_0$ for $a(T)$ and $b(T)$, as the effect is well reproduced [solid line in Fig.~\ref{fg:s6}(a)] by calculating the temperature dependence of the strain from the phenomenological fits to $a(T)$ and $b(T)$ (solid lines in Fig.~\ref{fg:abc}).  While $\eta(T)$ does not appear to require any significant contribution from the low-energy carriers, the temperature dependence of the strain might feed back on the electrons and contribute to the observed nematic response.

To compare the behavior of the thermal variation of the strain for our samples, we have plotted the change in strain, $\eta(T)-\eta({\rm10\,K})$, in Fig.~\ref{fg:s6}(b).  As one can see, there is remarkably similar behavior for the three underdoped samples.  The $x=0.93$ sample shows the largest change in strain, but with less upward curvature on cooling.  The change between $x=0.93$ and 0.75 is known from thermal expansion studies \cite{mein01}, but the lack of change in the underdoped region is intriguing.

\section{Discussion}
\label{sc:disc}

We have seen that the charge transfer from the chains to the planes in \ybco\ is screened by $c$-axis displacements of intervening ions.  Relaxation of these displacements with temperature implies a return of a fraction of the holes from the planes to the chains on warming.  The point here is that electronic behavior in the planes has a substantial coupling to $c$-axis displacements of atoms in and out of the planes.

Of course, we have only considered the average structure.  It is well known that, for underdoped orthorhombic YBCO crystals, the \och\ atoms order in filled and empty Cu-O chains, the periodicity of filled chains varying with the oxygen content \cite{beye89,zimm03}.   Though the chain ordering is never long-range, the periodicity is nevertheless indicated by diffuse superlattice peaks.   Analyses of the {\bf Q}-dependent superlattice intensities indicate that there are noticeable displacements of most atoms in the structure in response to the partial chain filling \cite{simo93b,gryb94,isla02,isla04}.  In particular, detailed studies of the structure in small crystals with $x=0.50$ found significant displacements of the Y and Ba ions along the $a$ axis, transverse to the chains, and of the \cupl\ and \oap\ atoms along the $c$ axis \cite{simo93b,gryb94}.

So far, we have only considered static variations of the average structure.  One may expect that there should also be significant dynamic coupling, and, indeed, there is good evidence for this.  For example, high-resolution EELS studies of \bscco\ indicate that the bosonic excitations to which electrons couple  correspond to $c$-axis polarized optical phonons \cite{qin10,vig15,phel93}.  Prominent peaks near 45~meV and 80~meV correspond to modes involving in-plane oxygens (bond-bending motion) and apical oxygens, respectively \cite{kaki96,tu02,kova04}. The importance of dynamic screening by $c$-axis phonons has also been demonstrated in an optical-pump--electron-probe study of a La$_2$CuO$_{4+\delta}$ film \cite{gedi07,rado08}.    In \ybco, the infrared-active modes of the apical oxygens vibrating along the $c$ axis are coupled to the electronic conductivity \cite{home95}.  In fact, Kaiser {\it et al.} \cite{kais14} found that driving the apical oxygens with a mid-infrared pulse dynamically increases the coherent $c$-axis conductivity.  At least part of this response should be a consequence of driving charge transfer between the planes and chains.

Given the substantial static and dynamic responses, one might expect significant local displacements in response to disorder.   In this context, it is interesting to consider the analysis of atomic displacements associated with the zero-field CDW in underdoped YBCO by Forgan {\it et al.} \cite{forg15}.  The biggest displacements are along the $c$ axis, involving Ba, \cupl, and, especially, the \opl\ sites.  The displacements have reflection symmetry about the chain layer, and the $L$ dependence of the scattering requires correlations only within one unit cell  above and below a chain.   An important observation is that the \opl\ displacements have a quadrupolar pattern, with sites along the $a$ and $b$ axes displaced in opposite directions along the $c$-axis \cite{forg15}.  Since the average structure of each CuO$_2$ plane is dimpled, this modulation decreases half of the \cupl-\opl\ bond lengths, and increases the other half.  One would expect the holes to prefer the short bonds, and hence this pattern reproduces the $d$ density form factor proposed by several theorists \cite{metl10,sach13,efet13,davi13} and confirmed by experimentalists \cite{fuji14b,comi15a}.

The symmetry about the chain layer and the short correlation lengths suggest that the modulation is driven by local charge transfer between a chain segment and neighboring bilayers.  That the chains might drive a modulation is plausible, as there is evidence from scanning tunneling microscopy (STM) measurements on \ybco\ for CDW modulations with wave vector of $\sim0.3b^*$ along the chains and a correlation length of a few lattice spacings in the transverse direction \cite{derr02,maki05}.  Nuclear magnetic resonance (NMR) studies of \cuch\ sites in filled chains indicate the development of an increased linewidth that has been interpreted as evidence of CDW order \cite{grev00} or Friedel oscillations \cite{yama06,mori04} in CuO chains.  [We note that an angle-resolved photoemission (ARPES) study by Zabolotnyy {\it et al.} \cite{zabo12} reported an absence of a gap in the chains of underdoped YBCO.  Of course, the ARPES measurements were performed in zero magnetic field, while the NMR studies used fields of 0 to 8~T.]

A new CDW order develops above a threshold magnetic field of 10~T applied along the $c$ axis for $T<T_c$ \cite{wu11,wu13a}.  X-ray diffraction measurements at high field find that, in addition to the low-field CDW signal, new peaks appear that are incommensurate only along the $b$ direction and that are commensurate along $c$, with a significantly enhanced correlation length \cite{gerb15,chan15}.  The observation that the new peak intensities for $(0,k,l)$ reflections are strong for $l=1$ but negligible for $l=0$ is consistent with a displacement pattern that is reflection symmetric about the mid-plane of a bilayer, rather than about a chain layer.

The low-field CDW is associated with an anomaly found in low-frequency phonon branches \cite{blac13b,leta14}.   For the high-field CDW, it may be relevant to consider the phonon anomaly that has been seen at ${\bf q}=(0,0.3)$ in the bond-bending branch of the CuO$_2$ planes in YBa$_2$Cu$_3$O$_7$ at low temperature \cite{raic11b}.  This branch connects to the $B_{1g}$ Raman-active mode at ${\bf q}=0$, which exhibits a softening below $T_c$ for samples near optimum doing \cite{alte92,rezn95}.  An important experimental feature is that the anomaly at finite {\bf q} occurs only along $b^*$, and not along $a^*$ \cite{raic11b}; an anomaly at the same wave vector, with the same anisotropy also occurs in the Cu-O bond-stretching mode \cite{pint05}. 

\begin{figure}[t]
\begin{center}
\includegraphics[width=0.9\columnwidth]{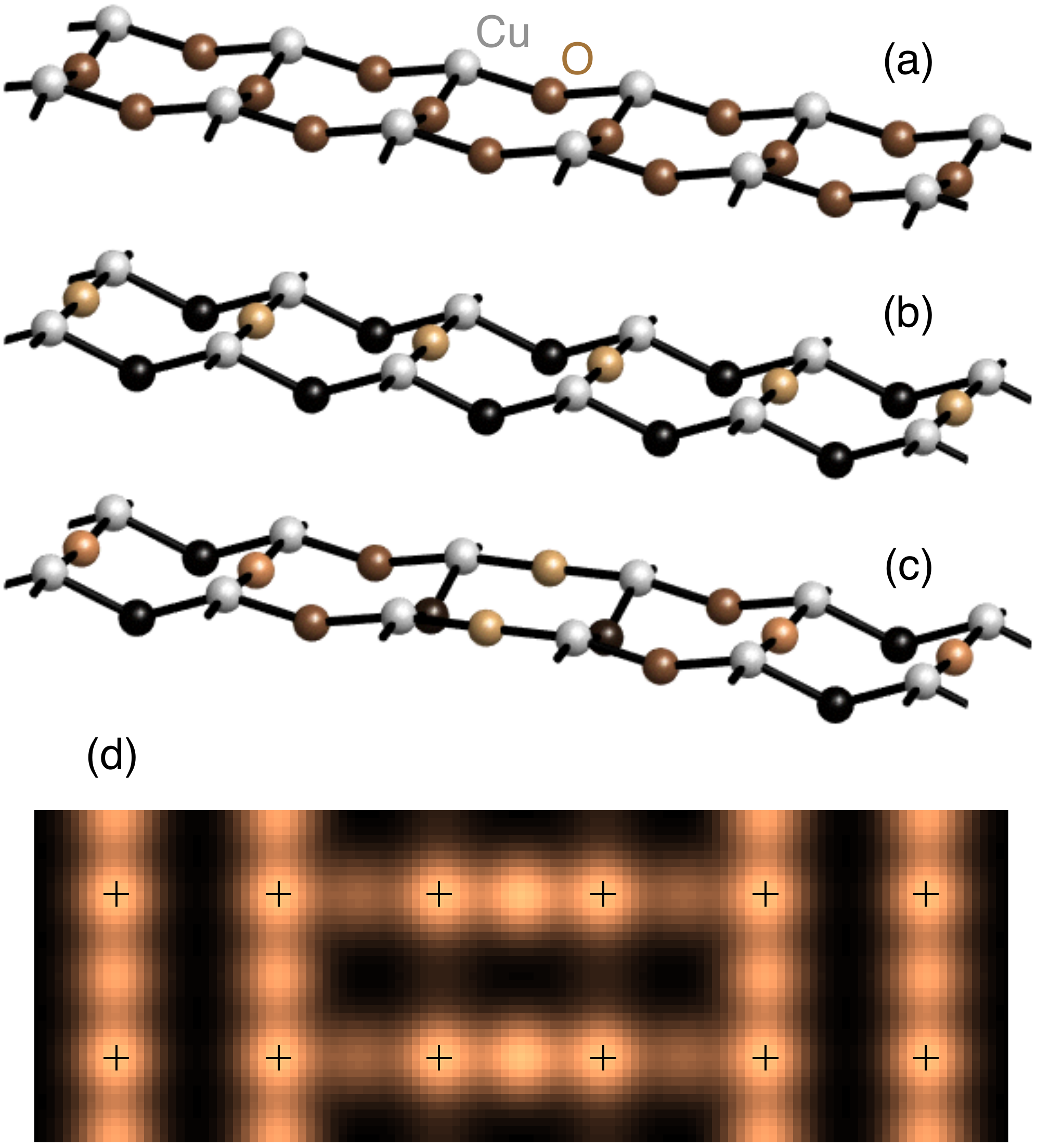}
\caption{\label{fg:stm} 
(a) Upper CuO$_2$ plane of a bilayer in \bscco, with O sites (brown) dimpled towards the Ca layer below, and Cu atoms (gray) above.  (b) Same layer after application of $B_{1g}$ symmetry displacements of the O sites, with O atoms shaded according to height along the $c$ axis, dark being furthest from the Cu layer.  (c)  Modulation of the $B_{1g}$ displacements at  ${\bf Q}=(0.25,0)$.  (d) Resulting pattern if we associate hole density with the O sites with the shortest bond lengths in (c).  Here, Cu sites ($+$) are also set to the same color as the hole-rich O sites, for comparison with STM images in \cite{kohs07}. }
\end{center}
\end{figure}

The CDW modulation in \bscco\ as detected by STM \cite{kohs07,dasi14} shares the $d$-wave form factor \cite{fuji14b,hami16} detected in YBCO \cite{comi15a}; however, it appears to be locally uni-directional in \bscco\ \cite{hami16}, similar to the high-field CDW in YBCO.  There are no CuO chains in \bscco, but there is plenty of disorder, and one may anticipate significant $c$-axis atomic displacements in response to that disorder.  Let us consider the impact of bond-bending displacements on the CuO$_2$ layers in \bscco.  In Fig.~\ref{fg:stm}, applying the $B_{1g}$ displacement pattern to the unperturbed, dimpled CuO$_2$ structure of (a) leads to the anisotropic structure shown in (b).  Modulating these distortions at the CDW wave vector of ${\bf q}\sim(0.25,0)$, one obtains the pattern shown in (c).  Associating hole density with the O sites along shortened bonds, one obtains the pattern displayed in (d); this looks quite similar to the images obtained in STM studies \cite{kohs07}.

It is interesting to compare with \lbco\ and \lsco.  As pointed out by Raghu {\it et al.} \cite{ragh12}, these compounds lack a charge reservoir layer.  As a consequence, the screening of Coulomb interactions is different.  The main instability of the planes is the tilting of the CuO$_6$ octahedra, which can result in low-temperature phases having long and short Cu-O bonds \cite{axe89,bozi15,jaco15}, as in the $B_{1g}$ modulations discussed above for \ybco\ and \bscco.   The bond-length anisotropy pins both charge and spin stripes \cite{huck11}.  The stripe order is compatible with pairing and superconductivity \cite{li07}, but it appears that the superconducting wave function forms a pair density wave \cite{berg09b,lee14,frad15}, which can frustrate interlayer Josephson coupling.

Given that the optimum doping concentration for charge order is $p \sim 0.12$ in all of the cuprate families studied \cite{huck11,park10,blan14,huck14,tabi14,comi15c}, it is reasonable to look for common features among these systems.  While the families other than LBCO and LSCO generally have significant spin gaps in the relevant doping range, there is evidence for electron nematic behavior \cite{lawl10,daou10} that couples to the local CDW order \cite{mesa11,hami15b}.  There is also photoemission evidence for some degree of pairing correlations in antinodal states at $T> T_c$ \cite{yang08,yang11,rebe12,kond13} defining the Fermi arcs \cite{norm98} or pockets \cite{yang11}.  If $c$-axis atomic displacements are present due to out-of-plane disorder, it may be reasonable for them to be modulated at a wave vector that couples the ends of the Fermi arcs, creating a gap between the less-correlated arc states and the pairing correlations in the antinodal region; indeed, a theoretical analysis has indicated that the bond-bending mode should couple to states with increasing strength as one moves from the nodal to the antinodal region \cite{deve04}.  The result might be a bosonic CDW.  (Think of a CDW along a charge stripe, formed from pairs, similar to the state described in \cite{chen04a}.)  Such a state would tend to compete somewhat with superconductivity, as the coherent pairing forms along the Fermi arcs \cite{lee07,vish12}.  We note that this picture differs from the PDW perspective of Lee \cite{lee14}, but is consistent with the concept that the CDWs in YBCO are not fundamental instabilities of the planes \cite{mish15,zhu15} but may be a consequence of incidental features \cite{mish15,hami16}.

\section{Summary}
\label{sc:sum}

We have considered how charge transfer from chains to planes in \ybco\ is screened by atomic displacements of oppositely-charged ions.  Neutron powder diffraction measurements show that such displacements are temperature dependent, implying a change in the charge transfer.  These results provide context for considering the CDW orders recently discovered in several cuprate families.  Displacements of in-plane O atoms along the $c$ axis, with patterns consistent with experimentally-observed phonon anomalies, are compatible with reported images of electronic modulations in \bscco.

\acknowledgments

We thank J. C. Davis, M. R. Norman, and W. G. Yin for helpful discussions.  This project was conceived and samples were prepared with support from the Center for Emergent Superconductivity, an Energy Frontier Research Center funded by Office of Basic Energy Sciences (BES), Division of Materials Sciences and Engineering, U.S. Department of Energy (DOE).  Work at Brookhaven was supported by the BES, US DOE, through Contract No.\ DE-SC00112704.  The experiment at ORNL's Spallation Neutron Source was sponsored by the Scientific User Facilities Division, BES, U.S. DOE.


%

\end{document}